\documentclass[10pt,journal,compsoc]{IEEEtran}
\usepackage[textsize=scriptsize]{todonotes}
 
\RequirePackage{soul}
\RequirePackage{url}
\usepackage[square,sort,comma,numbers, compress]{natbib}
\usepackage[yyyymmdd]{datetime}
\usepackage{amsmath}
\usepackage{amssymb}
\usepackage{amsthm}
\RequirePackage[inline]{enumitem} \setlist{nosep}
 
\def\framework{{\ttfamily Gamma\-Cas}} 
\def\shortname{\framework}

\ifCLASSINFOpdf
 
\else
  
\fi

\begin{document}
%
\title{Incomplete Gamma Integrals for Deep Cascade Prediction using Content, Network, and Exogenous Signals}

\author{Subhabrata Dutta, Shravika Mittal, Dipankar Das, Soumen Chakrabarti, and Tanmoy Chakraborty
\IEEEcompsocitemizethanks{\IEEEcompsocthanksitem S. Dutta is with the Dept. CSE, Jadavpur University, India. 

\IEEEcompsocthanksitem S. Mittal is presented with Adobe, India. The work was done when she was with IIIT Delhi, India. 

\IEEEcompsocthanksitem D. Das is with Dept. of CSE, Jadavpur University, India. 

\IEEEcompsocthanksitem S. Chakraborti is with Dept. of CSE, IIT Bombay, India

\IEEEcompsocthanksitem T. Chakraborty  is with Dept. of CSE, IIIT Delhi, India. 

}
}

\IEEEtitleabstractindextext{%
\begin{abstract}
The behavior of information cascades (such as retweets) has been modeled extensively. While point process-based generative models have long been in use for estimating cascade growths, deep learning has greatly enhanced diverse feature integration.  We observe two significant temporal signals in cascade data that have not been emphasized or reported to our knowledge.  First, the popularity of the cascade root is known to influence cascade size strongly; but the effect falls off rapidly with time.  Second, there is a measurable positive correlation between the novelty of the root content (with respect to a streaming external corpus) and the relative size of the resulting cascade.  Responding to these observations, we propose \shortname, a new cascade growth model as a parametric function of time, which combines deep influence signals from content (e.g., tweet text), network features (e.g., followers of the root user), and exogenous event sources (e.g., online news).  Specifically, our model processes these signals through a customized recurrent network, whose states then provide the parameters of the cascade rate function, which is integrated over time to predict the cascade size.  The network parameters are trained end-to-end using observed cascades.  \shortname\ outperforms seven recent and diverse baselines significantly on a large-scale dataset of retweet cascades coupled with time-aligned online news --- it beats the best baseline with $18.98\%$ increase in terms of Kendall's $\tau$ correlation and $35.63$ reduction in Mean Absolute Percentage Error.  Extensive ablation and case studies unearth interesting insights regarding retweet cascade dynamics.
\end{abstract}

}

\maketitle

\IEEEdisplaynontitleabstractindextext

%
\IEEEpeerreviewmaketitle

\section{Introduction}
\label{sec:Intro}

(Re)sharing is a common way in which content spreads in social networks.  A \emph{root user} posts some content (such as a photo or an article) and then \emph{friends} or \emph{followers} of that user share it with their friends, and so on, resulting in a \emph{cascade}.  In such a cascade tree, information flows from the root to the leaves.  In case of Twitter, resharing is called \emph{retweeting}.  The size, duration, and intensity of a reshare cascade are important indicators of user engagement at various levels: within the topic, the community, or the social media platform at large.  Modeling user engagement is useful in political discourse mining, market trend analysis, and user-persona detection.

Predicting the progression of a cascade, given early observations at its onset, is known to be a challenging problem \citep{salganik2006experimental, 10.1145/2566486.2567997, watts2012common, hofman2017prediction}.  Early approaches \citep{wang2018retweet, zhao2018attentional} relied on three types of features (network structure, root content, and initial observations along time) for modeling the growth of reply trees.  Self-exciting point processes \citep{DBLP:conf/icwsm/KobayashiL16, 10.1145/3038912.3052650} were also employed as generative models.  Recently, exogenous influence has been incorporated~\citep{jia2019jump, 10.1145/3394486.3403251}. Neural methods, particularly graph embedding-based techniques, are quickly becoming popular~\citep{DBLP:conf/www/LiMGM17,10.1145/3132847.3132973}.

Different existing approaches suffer from specific limitations. Extensive feature engineering provides remarkable performance over specific platforms. But they fail to generalize as importance and interdependence of different features vary sharply over different platforms. Pure point-process based models, however simple and explainable, do not take important signals of cascade growth (e.g, content-based features). They rely completely on the numerical growth of the cascade over the observed time to predict future behavior. Previous studies~\citep{10.1145/2983323.2983812} as well as our experiments suggest that the predictions of such models are often adversely affected by noise in the observed cascade. Prior neural models often heavily depend on the graph structure of the cascade growth. In most platforms, however, only the cascade participants are observable and not the exact cascade formation path (i.e, if a retweeter is a common follower of two previous retweeters, it is ambiguous to decide which one of them is the predecessor in the cascade graph). Moreover, most of these approaches do not model cascade growth as an explicit function of the prediction horizon. They need to be trained separately for predicting on different prediction horizons.

\begin{figure}[t]
\centering
\includegraphics[width=\columnwidth]{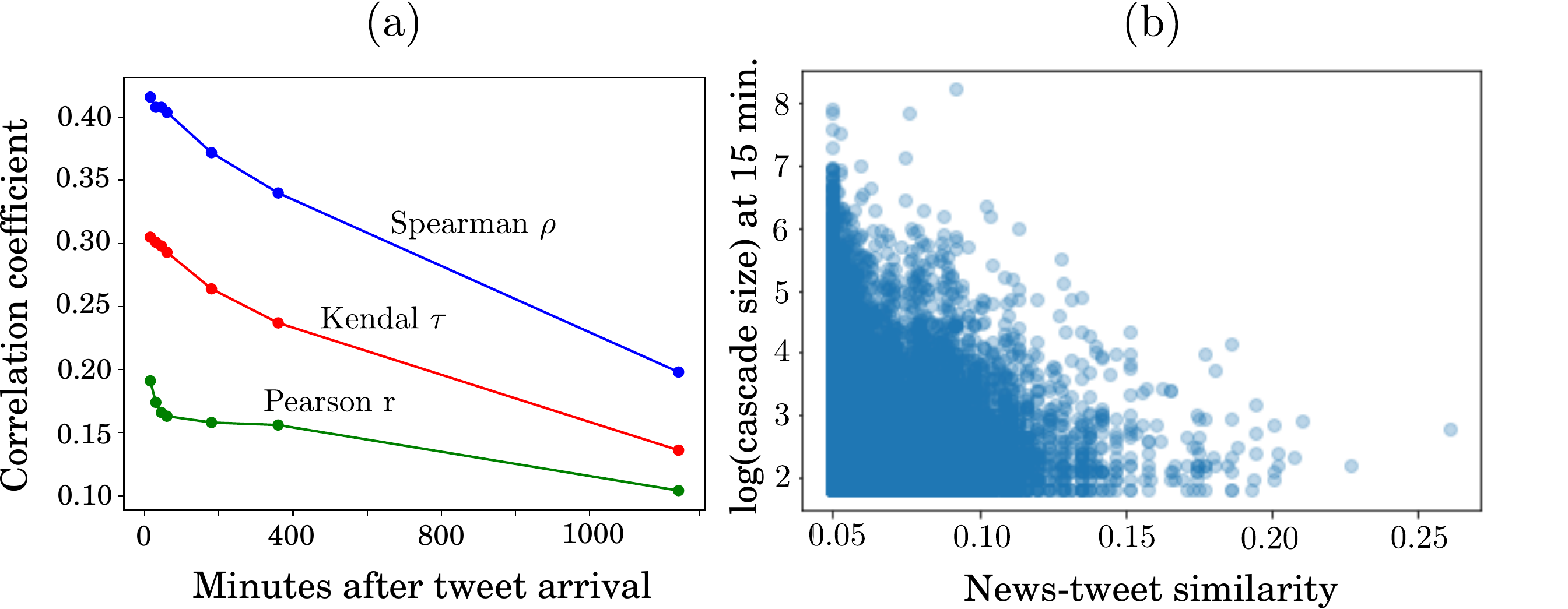}
\caption{(a) Correlation between root user's follower count and cascade size at different time after the arrival of the tweet. All three correlation coefficients indicate a decreasing influence of the root follower as the cascade grows further in time.
(b) The ``novelty premium'': tweets that are not mere repetition of current news enjoy greater cascade rates. Average unigram and bigram similarity between a tweet and the news articles published within $12$ hrs. before its arrival is plotted against the log of cascade growth (starting $15$ mins.). The later value signifies the virality of the tweet among its first responders. We observe a weakly negative correlation ($-0.09$ Spearman's $\rho$) but with $p$-value $<10^{-5}$.}
\label{fig:intro_fig}
\end{figure}

Our point of departure is the recognition of certain delicate temporal dynamics that existing cascade prediction methods seem unable to exploit, despite their rapidly increasing sophistication.  As an example, \figurename~\ref{fig:intro_fig}(a) shows that,  although the root user's popularity (follower count) is initially strongly predictive of cascade growth rate, the effect is not stationary, but rapidly fades with time. 
As another example, \figurename~\ref{fig:intro_fig}(b) shows a scatter of cascade sizes (logarithmic) achieved in 15 minutes against the content similarity between the root tweet and a body of news articles published shortly before and after the root tweet.  It hints at a certain ``novelty premium'' --- text that is not mere repetition of current news enjoys greater cascade rates.

Guided by observations like the ones narrated above, we present \textbf{\shortname}, a novel deep model for cascade prediction.  We directly model the gradient of cascade growth as a trainable neural function of content, network, and exogenous features.  Specifically, we monitor network (popularity) features evolving through time, and feed (continuous forms of) these features into a novel LSTM \cite{hochreiter1997long} variant, whose \emph{hidden states are then mapped to parameters that dictate the gradient of cascade growth}.  Textual and exogenous features modulate how LSTM states influence the temporal process parameters.

The gradient of cascade growth is then integrated over the past to predict the size of the cascade at a given time beyond the observation horizon.  Inspired by many natural growth processes~\cite{milbrandt2005multimoment,seifert2006two}, we model cascade trajectory as an incomplete gamma function by integrating its temporal derivative numerically. This allows us, during training, to back-propagate prediction errors and train all model weights end to end.



We report on extensive experiments using 342,111 resharing cascades from Twitter, temporally aligned with 206,180 news articles published online on 5,138 news sources.  We compare \shortname{} against several recent competitive approaches: a basic Hawkes process, SEISMIC \cite{10.1145/2783258.2783401}, TiDeH, a time-dependent Hawkes Process \citep{DBLP:conf/icwsm/KobayashiL16}, NeuralPointProcess~\citep{neuralnetwork-pointprocess}, CasPred \citep{10.1145/2566486.2567997}, DeepHawkes \citep{10.1145/3132847.3132973}, DeepCas~\citep{DBLP:conf/www/LiMGM17} and ChatterNet \citep{10.1145/3394486.3403251}.
\shortname{} achieves lower mean absolute percentage prediction error compared to these prior systems.  It is more stable and robust to variations in prediction horizons, compared to some prior systems.  Another benefit of \shortname's transparent network design is that, by correlating observable features against the parameters involved in the time integration, we get additional insights into the factors that govern cascade dynamics.

Summarizing, our major contributions are as follows: 
\begin{itemize}[leftmargin=*]
    \item We propose \framework, a novel framework  for reshare cascade prediction which incorporates content, network and exogenous signals over observable cascade progress to learn parametric representation of cascade growth at a future time. \framework\ achieves a Kendall's $\tau$ correlation of $\mathbf{0.63}$  ($\mathbf{25.06}$ Mean Absolute Percentage Error) between predicted and actual size of the cascade at $24$ hours after only $6$ hours of early observation.
    \item We collect and contribute a large-scale dataset of recent retweet cascades with a temporally aligned stream of online news articles.
    \item We compare \framework\ with several recent baselines for cascade size prediction developed upon generative, feature-driven, and neural network-based approaches. While \framework\ outperforms each of these baselines by a significant margin, we also investigate the behaviors of these baseline models on our dataset.
    \item We perform in-depth ablation and case study using \framework\ to investigate into the different signals influencing its parameter estimation. We present insights from these experiments which may be of independent interest.
\end{itemize}

\noindent
\textbf{Organization:} The rest of the paper is organized as follows:
\begin{itemize}
    \item We review the relevant literature on cascade and popularity prediction in Section~\ref{sec:related-work}, focusing on point-process and neural methods that incorporate different influence signals.
    \item \framework\ is presented in Section~\ref{sec:model} with detailed descriptions of its various functional components.
    \item In Section~\ref{sec:experiments} we describe the dataset preparation, training protocols of \framework, baseline methods and ablation variants of \framework.
    \item We present experimental results in Section~\ref{sec:results}.
    \item We conclude with important observations and possible future direction in Section~\ref{sec:End}.
\end{itemize}

\noindent
\textbf{Reproducibility:} To encourage reproducible research, we present detailed hyper-parameter configurations in Section~\ref{subsec:training}. Moreover, we  supplement our submission with dataset and source code of \shortname{}, available at: \url{https://github.com/LCS2-IIITD/GammaCas}.

\section{Related Work}
\label{sec:related-work}




Prior works in the field of information cascade modeling can be broadly distinguished into two categories: {\em Macro} cascade modeling focused on the overall growth and structural properties of a cascade (e.g., retweet count prediction) \citep{10.1145/2566486.2567997,DBLP:conf/icwsm/KobayashiL16} and {\em Micro} cascade modeling which investigates the behavior of individual agents participating in the cascade (e.g., retweeter prediction)~\citep{topo-lstm,deep-diffuse}. Our work specifically aligns with the macro category.

{\bf Feature-driven cascade modeling.} Among the earliest of works, \citet{10.1145/2566486.2567997} studied the structural and temporal properties of resharing cascades and came up with a feature-driven strategy to devise a classification problem: after observing a cascade reaching a size $k$, what is its probability of reaching size $nk$? \citet{10.1145/1935826.1935845} attempted to identify potential influencers in a feature-driven approach to predict information cascades. 
To explore richer feature set of cascade dynamics, \citet{10.1145/2908131.2908155} conceptualized cascades as information flow along forests as opposed to the usual tree structure. 
Most of the feature-driven approaches have revolved around temporal features \cite{DBLP:conf/icwsm/PetrovicOL11,10.1145/2566486.2567997}, structural and network features \cite{10.1145/1935826.1935845,DBLP:conf/icwsm/WengMA14}, user features \cite{10.1145/1935826.1935845, zaman2014bayesian} and content features \cite{10.1145/2396761.2398634}. While feature-based approaches have produced seminal insights regarding the dynamics of cascade growth, they require heavily curated manual feature engineering that are exclusively platform-dependent.

{\bf Generative models for cascade prediction.} An alternative emerging approach that has seen significant success involve generative models that perceive cascades as temporal event arrival sequences, generates random arrival sequences conditioned on certain parameters, and finally maximizes a chosen likelihood function between the observed and generated sequences \cite{10.1145/2783258.2783401, DBLP:conf/icwsm/KobayashiL16}. \citet{DBLP:conf/aaai/ShenWSB14} sought to model item popularity over complex networks using a Reinforced Poisson Process model. \citet{crane2008robust} described the view dynamics of YouTube as an epidemic modeled by a self-exciting Hawkes Process. Multiple studies reported using Hawkes Process or its modified variations to predict retweet cascade size \cite{10.1145/2783258.2783401, DBLP:conf/icwsm/KobayashiL16,10.1145/2740908.2742744,10.1145/3038912.3052650}. In a cross-platform setting, \citet{10.1145/3038912.3052650} used a Hawkes process to model popularity growth of content in one platform controlled by endorsement provided in other platforms.
\citet{10.1145/2983323.2983812} combined feature-driven approach with Hawkes process for popularity prediction. \citet{sir-hawkes} proposed a hybrid of epidemic and self-excitation models to analyse diffusion cascades. Although not often applied to cascade modeling, recent advances have been used to model more complex dynamics of temporal point processes using neural networks \citep{neural-hawkes,neuralnetwork-pointprocess}. Other than point-process models, a few others explored epidemic models~\citep{10.1145/2396761.2398634,esis}, Bass model~\cite{yan2016sth,8896027}, Survival Analysis~\cite{10.1109/WI-IAT.2010.209,9157926}, Jump Processes \citep{jia2019jump}, etc. Despite their explainable behavior and zero need for heavy feature engineering, generative models are susceptible to adverse influences from outliers \citep{10.1145/2983323.2983812} and found less powerful at making precise predictions~\cite{10.1145/3132847.3132973}.

{\bf Neural network based methods.} Recently, neural models have facilitated more powerful representations of two major components of cascade predictions: recurrent neural architectures can learn the complex temporal dynamics of early observation without constrained approximations~\cite{DBLP:conf/ijcai/WangSLGC17} and graph learning methods render the integration of complex structural properties to be seamless~\cite{DBLP:conf/www/LiMGM17}. In their proposed model DeepCas, \citet{DBLP:conf/www/LiMGM17} sought to learn the structural properties of observed retweet cascade using random walk embeddings of the cascade graph and aggregated the dynamics using gated recurrent units with attention. DeepHawkes was proposed by \citet{10.1145/2983323.2983812} to translate the explainable behavior of Hawkes Process into the representational superiority of neural networks to predict retweet and citation cascades. In the absence of explicit knowledge about a social or information network like Reddit, \citet{10.1145/3394486.3403251} proposed ChatterNet to model the growth of reply cascades; their model integrates exogenous and endogenous influence to learn textual representations of content using time-evolving convolution kernels and aggregates the observed cascade growth using LSTMs. One implementation challenge regarding most of these models is their lack of flexibility to migrate to different observation/prediction horizons without retraining. Moreover, in most of the cases, the superior representation power of neural network-based models is shadowed by the lack of explainability and the inability to produce actionable insights from the learned representations.

{\bf Exogenous influence over cascade growth.} While the mentioned works mostly focus on driving factors of cascade growth implicit to the cascade and the platform, signals exogenous to the platform determine the virality and popularity of content heavily~\cite{10.1145/2339530.2339540, 10.1145/3394486.3403251}. Prior works seeking to identify the influence of exogenous event arrivals have explored point process with self and external excitation to model observed event sequences~\cite{koyama2020statistical}.  \citet{10.1145/3178876.3186121} attempted to demarcate opinion diffusion in Twitter under the influence of exogenous influence from endogenous ones. \citet{DBLP:journals/jiis/BroxtonIVW13} investigated the influence of external information sources on virality of online video content. Cascade predictions based on cross domain influences are specialized scenarios of modeling and exploiting signals external to a platform, i.e, predicting YouTube view cascades from Twitter cascades~\cite{DBLP:journals/tmm/RoyMZL13}. 
\citet{10.1145/3394486.3403251} employed a similar strategy to incorporate exogenous signals; with the target domain being Reddit, their source domain of external influence was free-flowing new-streams on online news portals.

Given this vast prior development in modeling cascade dynamics, our proposed \framework\ model seeks to deliver a generalizable, flexible model for cascade growth prediction, similar to the generative family while incorporating the powerful representation capability of neural methods in an end-to-end fashion to capture the temporal, network-based, content-based and exogenous influences on the cascade growth. 

{\bf Differences between ChatterNet \cite{10.1145/3394486.3403251} and \framework.} Among the discussed models for cascade and popularity predictions, ChatterNet seeks to use a set of influence signals similar to ours. It predicts the future chatter intensity under a submission on Reddit, defined as the number of comments posted under that submission. However, there are some key differences as follows: (i)~Owing to the closed definition of Reddit's communities (i.e., subreddits), the original design of ChatterNet is able to characterize {\em endogenous influences} in terms of contemporary submissions posted within that subreddit. This is not at all possible for a Twitter-like open platform. Instead, \framework\ uses the social network information (i.e, follower count of users) to model the endogenous influence. ChatterNet is not developed to handle such information because Reddit does not provide any. (ii)~Being a purely deep learning based model like DeepCas~\citep{DBLP:conf/www/LiMGM17}, ChatterNet does not learn the prediction function as explicitly dependent on the prediction horizon. Therefore, a new training setup is needed for each different prediction horizon. \framework\ overcomes this lack of flexibility by learning a parametric estimation of retweet arrival intensity and then performing numerical integration of the said intensity function over the prediction horizon. This novel hybrid of deep feature learning with numerical function approximation empowers \framework\ with the flexibility that, once trained, it may predict for arbitrary prediction horizons.

\section{Proposed Model}
\label{sec:model}

In this section, we describe \shortname{} in detail.  It has many modules which may appear complex, but we will justify their utility through ablation in Section~\ref{sec:experiments}.

\begin{figure*}
    \centering
    \includegraphics[width=0.9\textwidth]{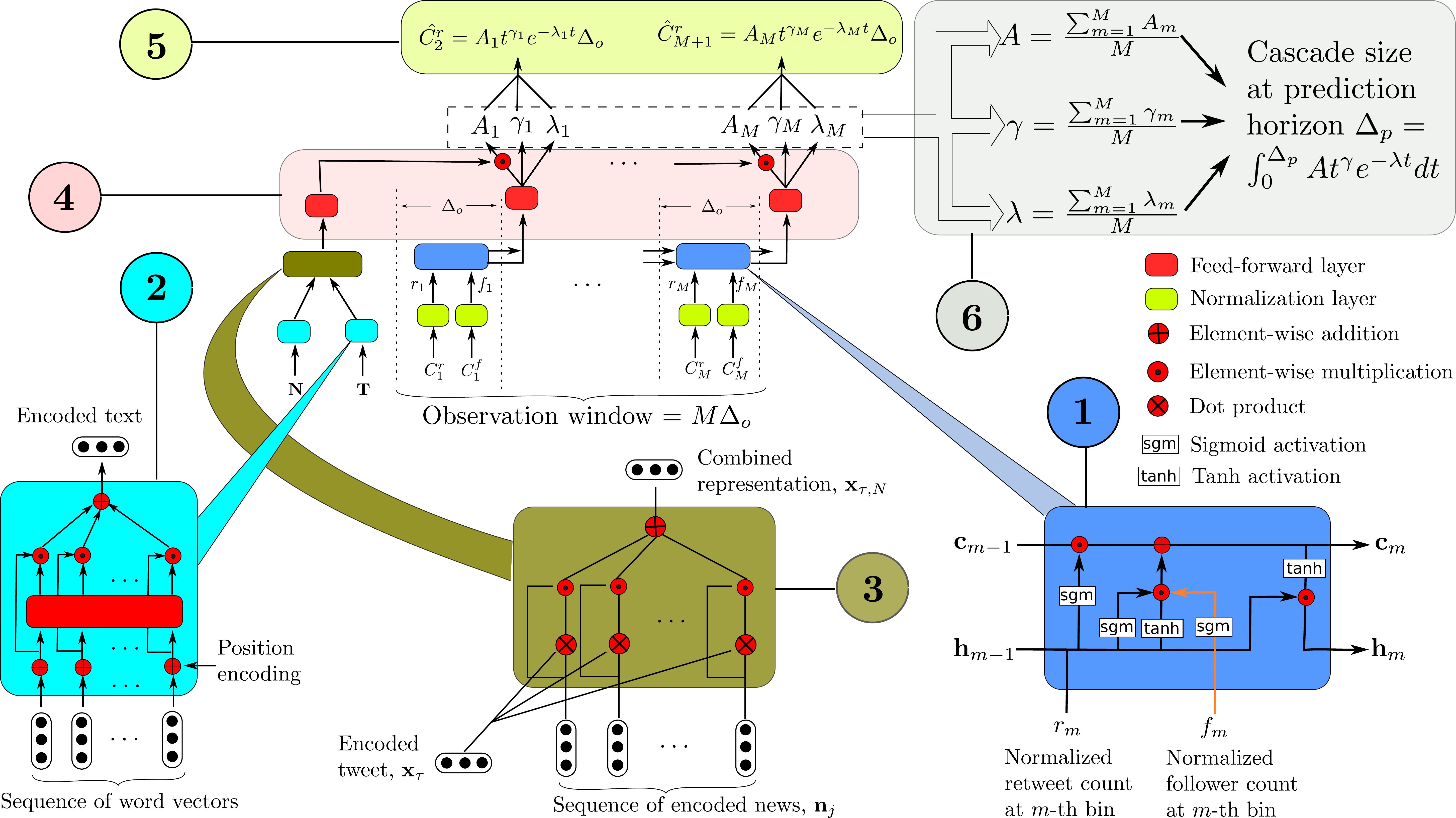}
    \caption{Design of \framework\ explained with its different modules. Retweet count and aggregate follower count at each observation bin ($\Delta_o$) is normalized and fed to (1) the modified LSTM layer (Section~\ref{subsec:lstm_retweet_arrival}). Textual content from tweet and news are processed in (2) the text processing module (Section~\ref{subsec:text_processing}) which performs word-wise attention and aggregation to generate a single vector per piece of text. Encoded tweet and sequence of news articles are then combined into a single representation in (3) a scaled dot-product attention layer (Section~\ref{subsec:news-tweet_attention}). Hidden state output from (1) at each bin and the news-tweet combined representation from (3) are then used in (4) the parameter estimation module to compute the parameters $A_m$, $\gamma_m$, and $\lambda_m$ for each bin $m$ (Section~\ref{subsec:param_compute}). In (5) the autoregressive module, the $m$-th set of parameters is used to predict the retweet arrival at $(m+1)$-th bin and the average-pooled parameters are used in (6) where the future cascade size at prediction horizon $\Delta_p$ is computed (Section~\ref{subsec:prediction}).}
    \label{fig:architecture}
\end{figure*}

\subsection{Preliminaries and problem definition} 
\label{subsec:prelims}
Let $\mathcal{G} = \{\mathcal{U}, \mathcal{E}\}$ be a directed graph representing the social network of Twitter, where $\mathcal{U}$ is the set of vertices representing the users and $e_{ij}\in \mathcal{E}$ if $u_j$ {\em follows} $u_i$ for any $u_i, u_j \in \mathcal{U}$. Therefore, the follower count of any given user $u_i$ translates to the out-degree of the corresponding node in $\mathcal{G}$.

Given a tweet $\tau$ posted by a user $u$ at time $t_0$, its {\bf retweet cascade at time $t>t_0$} can be defined as {\em an ordered sequence of retweet arrival timestamps along with the corresponding retweeter},  $\mathcal{R}^\tau_t = \{(t_i, u_i)|t_i>t_j\text{ for }i>j, t_i\leq t\}$. The exogenous event signals within any time frame $[t, t+\Delta t]$ are substantiated as the sequence of news articles $N(t, t+\Delta t):=\{(n_j, t_j)|t\leq t_j < t+\Delta t\}$, where $n_j$ is an article published at $t_j$.

For any given cascade $\mathcal{R}^\tau_t$, we define the {\bf early observation window} to be $(t_0, t_0+\Delta_{obs}]$. A model would estimate the future growth of the cascade upon observing the dynamics within this observation period. We also define a {\bf prediction horizon} $[t_0, t_0+\Delta_p], \Delta_p>\Delta_{obs}$, so that our problem translates to predicting $|\mathcal{R}^\tau_{t_0+\Delta_p}|$ upon observing $\mathcal{R}^\tau_{\Delta_{obs}}$, $\tau$, and $N(t_0-\Delta_{obs}, t_0+\Delta_{obs})$. Henceforth, for the sake of brevity, we will consider $t_0=0$ in general. 

{\bf Notation.} Table \ref{tab:not_denot} summarizes important notations and denotations. While describing \shortname, we use bold lower cased symbols to denote vector inputs and outputs, and bold upper cased symbols to denote sequences of vectors as well as the trainable parameters of \shortname. 

\subsection{Parametric estimation of cascade growth}
As \citet{10.1145/2783258.2783401} suggested, such a cascade can be either in a {\em supercritical} stage (rate of cascade growth is increasing) or in a {\em subcritical} stage (rate of cascade growth is decreasing) at different points of time, depending on multiple factors like the relevance of the content expressed by the piece of tweet, out-degree of the nodes participated in the cascade by that time, inter-arrival time of retweets, etc. Extending discrete-valued $\mathcal{R}^\tau_t$ to a continuous, real-valued map of time, we can redefine these two stages as $\frac{d^2 |\mathcal{R}^\tau_t|}{d t^2} \geq 0$ (supercritical) or $\frac{d^2 |\mathcal{R}^\tau_t|}{d t^2} < 0$ (subcritical).  Such a rate of growth can be modeled as a simple product of two functions of time,
\begin{equation}
\small
\label{eq:cascade_growth}
    \frac{d |\mathcal{R}^\tau_t|}{d t} = \Psi_1(t)\Psi_2(t)
\end{equation}
constrained with the following conditions: i) $\Psi_1(t), \Psi_2(t)>0$, ii) $\frac{d \Psi_1}{d t}>0, \frac{d \Psi_2}{d t}<0$ and iii) $\lim_{t \to +\infty}\Psi_1(t)\Psi_2(t) = 0$. The first condition ensures a monotonous growth of the cascade, while the second and third conditions ensure a possible initial supercritical growth followed by a mandatory subcritical growth. 

Simple choices for such functions would be a polynomial $\Psi_1$ and an exponentially decaying $\Psi_2$. Concretely, we can approximate Eq.~\ref{eq:cascade_growth} in a parametric form as follows:
\begin{equation}
\label{eq:param_function}
    \frac{d |\mathcal{R}^\tau_t|}{d t} = At^\gamma e^{-\lambda t}
\end{equation}
where $A$, $\gamma$, and $\lambda$ are arbitrary constants.

The choice of such a function restrains $\frac{d |\mathcal{R}^\tau_t|}{d t}$ to a single ``hill''-shaped curve corresponding to a single supercritical and single subcritical phase, whereas real cascades may have multiple consecutive super- and subcritical phases. The growth rate of such cascades can be easily approximated as:
\begin{equation}
\small
    \frac{d |\mathcal{R}^\tau_t|}{d t} = \sum_k A_k(t-\phi_k)^{\gamma_k} e^{-\lambda_k (t-\phi_k)}
\end{equation}
where $A_k, \lambda_k, \gamma_k$ correspond to the growth parameters of the $k$-th "hill" in the cascade growth and $\phi_k$ correspond to its starting time.

\begin{table}[!t]
    \centering
    \caption{Denotation of important notations used.}
     \label{tab:not_denot}
    \begin{tabular}{c|c}
    \hline
        {\bf Notation} & {\bf Denotation} \\
    \hline
        $\mathcal{R}^\tau_t$ & Retweet cascade of tweet $\tau$ through time $t$\\
        $\Delta_{obs}$ & Intial observation window of cascade\\
        $\Delta_p$ & Prediction horizon for future cascade\\
        $\Delta_o$ & Binning size of the observation window\\
         $M$ & Number of bins in observation window\\
        $N(t_1, t_2)$ & News articles published within [$t_1$, $t_2$]\\
        $C^r_m$ & Total retweets within $m$-th observation bin\\
        $C^f_m$ & Total followers within $m$-th observation bin\\
    \hline
     \end{tabular}
     \vspace{-3mm}
 \end{table}

However, we are interested in the size of the cascade after a finite amount of time $T$, which is given by
\begin{equation}
\small
    \begin{split}
    \mathcal{R}^\tau_T &= \int_0^T \sum_k A_k(t-\phi_k)^{\gamma_k} e^{-\lambda_k (t-\phi_k)}dt\\
    &= \sum_k A_k \int_0^T (t-\phi_k)^\gamma_ke^{-\lambda_k (t-\phi_k)}dt\\
    &= \left. \sum_k \frac{A_k}{\lambda_k^{(\gamma+1)}}(-\Gamma(\gamma_k+1, \lambda(t-\phi_k))) \right|_{t=0}^{t=T}\\
    &= \sum_k \frac{A_k}{\lambda_k^{(1+\gamma_k)}}(\Gamma(\gamma_k+1, -\lambda_k\phi_k)-\Gamma(\gamma_k+1, \lambda_k(T-\phi_k)))
    \end{split}
 \end{equation}
where $\Gamma(s, z)$ are \textit{\textbf{incomplete Gamma functions}}. Since any generalized incomplete Gamma function can be represented as a finite sum of modified Bessel functions of the first kind~\cite{veling2011generalized}, the above form is equivalent to a single Gamma function with suitably chosen values of the arbitrary constants. Therefore, we choose to model cascade growth as a parametric function 
\begin{equation}
\small
\label{eq:model_equation}
    |\mathcal{R}^\tau_{\Delta_p}| = \int_0^{\Delta_p} At^\gamma e^{-\lambda t}
\end{equation}
where the parameters, in turn, are estimated as (neural) functions of $\mathcal{R}^\tau_{\Delta_{obs}}$, $\tau$, and $N(t_0, t_0+\Delta_{obs})$.

\subsection{Capturing temporal dynamics of retweet arrival}
\label{subsec:lstm_retweet_arrival}
At any time $t$, the rate of cascade growth $\frac{d |\mathcal{R}^\tau_t|}{d t}$ directly depends on the retweets arriving within $(t, t+\Delta)$ interval. The exact number of retweets (we denote it as $C^r_{t, t+\Delta}$) arrived within this interval directly adds to the size of the cascade. Moreover, each of the new retweets expands the number of potential future retweeters (i.e., susceptible nodes) by the out-degree of the current retweeter. 

To capture this temporal dynamics within the early observation phase, we quantize the observation window into $M$ consecutive, equal-sized bins of size $\Delta_o$ (i.e., $\Delta_{obs}=M\Delta_o$), where $M$ is an application-driven hyperparameter. We denote the total number of retweets arrived within the $m$-th bin as $C^r_m$, where $m\in [M]$. We aggregate the additional amount of susceptible nodes created within the $m$-th bin as $C^f_m = \sum_j outdegree(u_j),\ \forall (t_j, u_j)\in \mathcal{R}^\tau_{m\Delta_o}/\mathcal{R}^\tau_{(m-1)\Delta_o}$. 
Furthermore, as shown in Figure~\ref{fig:architecture} (5), we apply trainable normalization on the integer elements of $C^r_m$ and $C^f_m$ to avoid gradient saturation in the subsequent layers of our framework. The resulting values are denotes as $r_m$ and $f_m$, respectively.

The sequences $\{r_m\}_{m=1}^M$ and $\{f_m\}_{m=1}^M$ represent the temporal dynamics of cascade growth within the observation window, and a simple choice of architecture to model it would be from the Recurrent Neural Network (RNN) family. While LSTMs have been successfully applied to model temporal dependencies over long sequences, we modify information flow along the LSTM gates according to the intuitive knowledge of the retweet arrival dynamics. As shown in Figure~\ref{fig:architecture} (1), the modified LSTM cell in our architecture instantiates the following six operations:
\begin{equation}
\small
\label{eq:forget_gate}
    \mathbf{x}_g = \sigma(\mathbf{W}_g[r_m:\mathbf{h}_{m-1}] + \mathbf{B}_g)
\end{equation}
\begin{equation}
\small
\label{eq:input_gate}
    \mathbf{x}_{in} = \sigma(\mathbf{W}_{in}[r_m:\mathbf{h}_{m-1}] + \mathbf{B}_{in})
\end{equation}
\begin{equation}
\small
\label{eq:cell_gate}
    \mathbf{x}_c = \tanh(\mathbf{W}_c[r_m:\mathbf{h}_{m-1}] + \mathbf{B}_c)
\end{equation}
\begin{equation}
\small
\label{eq:follower_gate}
    \mathbf{x}_f = \sigma(\mathbf{W}_ff_m + \mathbf{B}_f)
\end{equation}
\begin{equation}
\small
\label{eq:cell_update}
    \mathbf{c}_m = \mathbf{c}_{m-1}\odot \mathbf{x}_g + \mathbf{x}_{in} \odot \mathbf{x}_c \odot \mathbf{x}_f
\end{equation}
\begin{equation}
\small
\label{eq:out_gate}
    \mathbf{h}_m = \mathbf{h}_{m-1}\odot \tanh(\mathbf{W}_h\mathbf{c}_m + \mathbf{B}_h)
\end{equation}
where $[:]$ denotes concatenation; $\odot$ denotes the Hadamard product; $\sigma$ signifies the sigmoid non-linearity; $\mathbf{c}_m$ and $\mathbf{h}_m$ correspond to the cell and hidden state of the LSTM after the $m$-th timestep (observation bin) respectively; $\mathbf{W}_g$, $\mathbf{W}_{in}$, $\mathbf{W}_c$, $\mathbf{W}_f$, $\mathbf{W}_h$ are the learnable weight matrices, and $\mathbf{B}_g$, $\mathbf{B}_{in}$, $\mathbf{B}_c$, $\mathbf{B}_f$, $\textbf{B}_f$ are the learnable bias matrices.

Equations \ref{eq:forget_gate}, \ref{eq:input_gate}, \ref{eq:cell_gate} and \ref{eq:out_gate} correspond to the operations performed by the original LSTM cell. However, Equation~\ref{eq:follower_gate} generates a modulation signal $\mathbf{x}_f$ from the out-degree of the participating user nodes at that step to control the contribution of their retweets. Equation~\ref{eq:cell_update} takes this modulation into account to update the cell state for the current step. Moreover, this modification decreases the size of the parameter space compared to the original LSTM. Assuming the state size to be $s$, the four weight matrices of the original LSTM cell would incur a total of $12\times s$ number of weight and bias parameters, while the modified one uses $10\times s$ parameters due to split inputs. 

\subsection{Processing textual content}
\label{subsec:text_processing}

We take every piece of text (tweet or news) as a sequence of words and compute a single vector representation of the text relevant to the downstream task, as shown in Figure~\ref{fig:architecture} (2).

We use a trainable embedding layer to map each word $w_i$ to a $d$-dimensional vector $\mathbf{v}_i\in \mathbb{R}^d$, converting a piece of text into a sequence of vectors $\mathbf{V}$. Typical content-sharing platforms like Twitter incur heavy traffic, with millions of textual pieces arriving each second. To speed up the processing, we intend to maintain parallel operations on $\mathbf{V}$. Consequently, we do not use any sequential architecture involving variants of RNN to encode the representation. Instead, we compute {\em positional encoding} vector~
\cite{DBLP:conf/nips/VaswaniSPUJGKP17} 
$\mathbf{p}_i\in \mathbb{R}^d$ as 
\begin{equation*}
    p^{(j)}_i = 
    \begin{cases}
      \sin(\omega_i) \text{ if $j$ is even} \\
      \cos(\omega_i) \text{ otherwise }
    \end{cases}
\end{equation*}
where $i,j\in \mathbb{N}$, $\omega_k = L^{-\frac{2k}{d}}$, $L$ is the maximum length of the input text sequence in the corpus, and $p^{(j)}_i$ denotes the $j$-th element of the vector $\mathbf{p}_i$. The embedded sequence of words, $\mathbf{V}$ is then transformed to a position encoded sequence $\mathbf{V}' = \{\mathbf{v}'_i|\mathbf{v}'_i=\mathbf{v}_i+\mathbf{p}_i\}$.

Next, for every token position, we compute an attention weight $\alpha_i$ using a feed-forward layer followed by a softmax activation:
\begin{equation}
\small
\label{eq:text_attention}
        \alpha_i = \frac{e^{s_i}}{\sum_i e^{s_i}}
\end{equation}
where $s_i = \mathbf{W}_a\mathbf{v}_i + \mathbf{B}_a$, $\mathbf{W}_a$ and $\mathbf{B}_a$ are learnable weight and bias matrices, respectively. We compute the final representation of the text as weighted aggregation of $\mathbf{V}'$ as $\sum_i \alpha_i\mathbf{v}'_i$. Intuitively, Equation~\ref{eq:text_attention} generates a word-wise attention weight sequence, which modulates the contribution of each word in the final representation of the text. 

We also experimented with more complex text encoding methods like Transformer encoder, Bi-LSTM encoder, and BERT. These models incurred higher training/inference cost in terms of memory and time with no significant improvement over our proposed method. As \citet{10.1145/3394486.3403251} suggested, popularity of a content in social media is majorly governed by simpler textual features like topic, polarity, etc. which can be easily captured by simpler models, and sophisticated NLP methods tend to be overkill. Furthermore, the per-word weights, $\alpha_i$, computed by this proposed approach further serve to explain the effects of the textual content of the tweet on the growth of the resulting retweet cascade.

\subsection{News-tweet attention as exogenous influence}
\label{subsec:news-tweet_attention}

For a given tweet $\tau$ and a sequence of news $N$, the text processing module outputs a single vector $\mathbf{x}_\tau$ and a sequence of vectors $\{\mathbf{n}_{j}\}$, respectively. As exogenous influence on cascade growth varies for tweets expressing different topics, we amalgamate the two signals to compute the final influence, as shown in Figure~\ref{fig:architecture} (3).

We compute an attention weight between the tweet representation $\mathbf{x}_\tau$ and a news representation $\mathbf{n}_j$ as
\begin{equation}
\small
\beta_{\tau, j} =
\operatorname{softmax}_j\left(\frac{\mathbf{x}_\tau^\top \mathbf{n}_j }{\sqrt{d}}\right)
\end{equation}
The scaling component $d^{-0.5}$ reduces the chance of $\operatorname{softmax}(\cdot)$ reaching saturation. Similar to the text processing module, the final representation of the exogenous influenced tweet text is computed as $\mathbf{x}_{\tau, N} = \sum_j \beta_{\tau, j}\mathbf{n}_{j}$. 

\subsection{Computing cascade growth parameters}
\label{subsec:param_compute}

The cascade growth parameters $A$, $\gamma$, and $\lambda$ (see Equation \ref{eq:model_equation}) are computed from the textual representation $\mathbf{x}_{\tau, N}$ and the observed cascade dynamics encoded by the modified LSTM, $\mathbf{h}_m$ (see Equation \ref{eq:out_gate}). We hypothesize that while the growth and decay parameters, $\gamma$ and $\lambda$, can be estimated from observing the retweet arrivals exclusively, the scaling parameter $A$ is dependent on the tweet text and the exogenous influence. 

We map $\mathbf{h}_i$ to three separate non-negative scalars, $A'_m$, $\gamma_m$ and $\lambda_m$, using three parallel feed-forward layers as follows:
\begin{equation}
\small
\label{eq:A_t}
    A'_m = \operatorname{relu}(\mathbf{W}_A\mathbf{h}_m + \mathbf{B}_A)
\end{equation}
\begin{equation}
\small
\label{eq:gamma_t}
    \gamma_m = \operatorname{relu}(\mathbf{W}_{\gamma}\mathbf{h}_m + \mathbf{B}_\gamma)
\end{equation}
\begin{equation}
\small
\label{eq:lambda_t}
    \lambda_m = \operatorname{softplus}(\mathbf{W}_{\lambda}\mathbf{h}_m + \mathbf{B}_\lambda)
\end{equation}
We choose these activations experimentally. While $\operatorname{relu}(\cdot)$ is the most straightforward activation function to ensure non-negative output, \framework\ suffers from the zero-gradient problem of ReLU while computing $\lambda_i$.

Next, we compute a modulation parameter emerging from the tweet and the exogenous signals as another non-negative scalar value and scale $A'_i$ as follows:
\begin{equation}
\small
    A_m = A'_m\operatorname{relu}(\mathbf{W}_\mu \mathbf{x}_{\tau, N} + \mathbf{B}_\mu) 
\end{equation}
where $W_\mu$ and $B_\mu$ are learnable parameters of a feed-forward layer.

\subsection{Final prediction}
\label{subsec:prediction}
From Equations \ref{eq:A_t}, \ref{eq:gamma_t}, and \ref{eq:lambda_t}, we estimate the cascade growth parameters for each observation bin. We apply average-pooling from these three sequences to get the cascade size parameters $A$, $\gamma$, and $\lambda$. For a given prediction horizon $\Delta_p$, the predicted size of the cascade can then be found by solving the integration in Equation~\ref{eq:model_equation}. We use $4$-th order Runge-Kutta method with fixed number of steps to solve this integration numerically and predict the cascade size at $\Delta_p$ as $Y_{\Delta_p}$.

Learning to estimate the aggregate parameters of cascade growth at some prediction horizon is the primary task which \shortname\ is designed for. However, within the observation window, a  fine-grained prediction modeling of retweet arrival is supposed to help the model learn more robustly.
We use a joint learning strategy in an autoregressive setting. At the $m$-th observation bin, we have already estimated the parameters $A_m$, $\gamma_m$, and $\lambda_m$. From these, we predict the aggregate retweet arrival at the $(m+1)$-th bin as $\hat{C}^r_{m+1} = A_mt^{\gamma_m}e^{-\lambda_m t}\Delta_o$.  The gradient from the loss  can be back-propagated through the quadrature~\citep{wehenkel2019unconstrained} as mentioned below.

{\bfseries Loss/cost function.} We use two different loss functions to train the model in the joint learning setting. As future cascade size varies largely, we use the {\bf Mean Absolute Percentage Error} between the predicted and actual cascade size at a prediction horizon $\Delta_p$, as suggested by \citet{10.1145/3394486.3403251}. For the autoregressive task of predicting retweet arrival in the next observation window, we use Mean Squared Error loss. The final loss function therefore becomes:
\begin{equation}
\small
\label{eq:loss-func}
    J = \frac{\big| |\mathcal{R}^\tau_{\Delta_p}| - Y_{\Delta_p}\big|}{|\mathcal{R}^\tau_{\Delta_p}|} + \zeta\sum_{m=1}^M(C^r_{m+1}-\hat{C}^r_{m+1})^2/M
\end{equation}
where $\zeta<1$ is a hyperparameter to set the relative importance of the autoregressive gradient.

\section{Experimental Setup}
\label{sec:experiments}
In this section, we present the dataset used in the experiments, the baselines and ablation variants of \framework\ considered for the comparison.
\subsection{Dataset}
\label{subsec:dataset}
As collecting retweet information and parallel news articles for existing datasets often result in lots of missing information, we proceed with curating a dataset of our own. Overall, we use a total of 239,478 and 102,633 retweet cascades, respectively, for training and testing purposes. To encode exogenous signal, we use a total of 206,180 news articles published online within the same time period as the cascades.

\begin{figure}[!t]
     \centering
    \includegraphics[width=0.9\columnwidth]{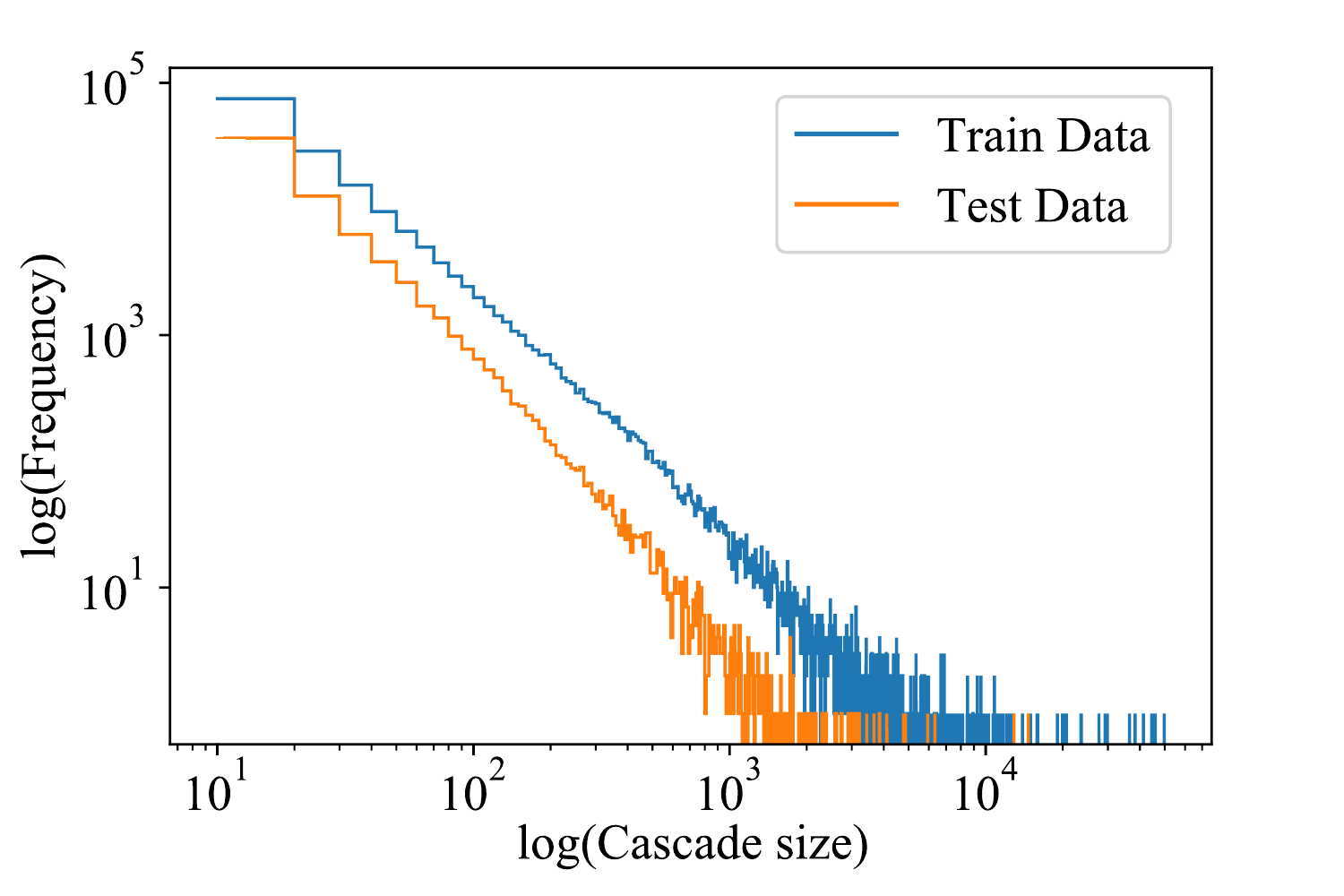}
     \caption{Log-log distribution of cascade sizes in training and test sets. A power law distribution of cascade size is maintained in both the sets.}
     \label{fig:data_distribution}
     \vspace{-0.5cm}
 \end{figure}

 Collecting retweet information from Twitter is cumbersome as the official API only returns $200$ recent retweets given a tweet id. We used their streaming API to collect tweets posted in real-time which allows us to collect retweets as independent entities. After collecting bulk tweet information for a $6$ month-long period, we map the retweets to their parent tweets and construct the cascade data. We considered only those tweets as cascade roots for which we have at least $15$ days of subsequent tweeting information after its posting. After discarding cascades with less than $10$ retweets, we finally end up with a total of 342,111 retweet cascades from \formatdate{25}{9}{2019} to \formatdate{25}{4}{2020}. After a random 70:30 train-test split, the training and test set contains 239,478 and 102,633 cascades, respectively. In Figure~\ref{fig:data_distribution}, we plot the cascade size distribution in both training and test sets and observe that both of them follow a power-law distribution.

 Aligned with the timeline of the retweet cascades, we crawled news articles published on the Web using the News-please crawler~\cite{Hamborg2017}. After discarding non-English news articles and news from sources that have less than $10$ articles published within this timeline, we end up with a total of 206,180 news articles from 5,138 different online sources.

 \subsection{Training protocols}
 \label{subsec:training}

While training \framework, we set the observation window $\Delta_{obs}$ to $6$ hours. with bin size of $5$ minutes. This results in an observation sequence (as well as LSTM time steps) of length $72$. To force \framework\ to learn the parametric representation of cascade growth based on predictions made at different prediction horizons, all predictions are made at $\Delta_p=12, 18, 24, 36, 48, 72, 120, 240, 360$ hours. 

 We use the headlines of news to represent the exogenous signal. After cleaning and tokenization, we set a maximum length for tweet and news headline texts to be $30$ and $36$, respectively. We set the word embedding dimension $d$ to be $256$. To initialize the embedding layer (see Section~\ref{subsec:text_processing}), we use pre-trained word vectors which we train on the tweets and news corpus jointly using Word2Vec~\cite{NIPS2013_5021}. The state-size of the modified LSTM layer (Section~\ref{subsec:lstm_retweet_arrival}) is set as $16$.

 We set the parameter $\zeta$ in Equation~\ref{eq:loss-func} experimentally on the development set; varying it from $0.1$ to $0.6$ with a step size of $0.05$, we found the best configuration to be $\zeta=0.25$.

 For mini-batch training of \framework, we use a batch size of $256$ (after varying it from $64$ to $512$ with a step size of $64$). \framework\ is optimized using the Adam optimizer with a learning rate $0.0025$ (optimal value found between the search range $0.0005-0.005$ with step $0.0005$). We trained \framework\ for a total of $40$ epochs. All of the mentioned hyperparameter search was done using $10\%$ of the training data as a development set. We use MAPE loss for predicting cascade size at $24$ hrs. prediction horizon as the search criteria.

 \framework\ is implemented using Tensorflow v2.0.0-beta0 on a Intel Xeon Silver $4116$ $2.10$GHz CPU with $48$ cores and $64$ GB RAM.

\subsection{Baseline methods}
\label{baselines}

To compare the performance of \framework, we implement a diverse set of baselines from generative, feature-based, and neural network-based families of frameworks.

\subsubsection{Generative baselines} We implement the following three self-excitation process-based models:

{\bfseries Hawkes.} We implement a univariate Hawkes Process-based model with the exponential kernel, optimized using maximum log-likelihood estimation to provide a basic generative baseline for future cascade size prediction on our data.

{\bfseries SEISMIC}, proposed by \citet{10.1145/2783258.2783401}, uses a self-exciting point process for retweet cascade prediction combined with the exposure provided by a user's follower base.

{\bfseries TiDeH}, a time-dependent Hawkes Process \cite{DBLP:conf/icwsm/KobayashiL16},  looks at how a cascade evolves with time considering the network structure and aging of information.

\subsubsection{Feature-driven baseline}
Following the work of \citet{10.1145/2566486.2567997}, 
we implement {\bf CasPred} to predict whether a given cascade will reach a particular size, exploiting rich, hand-crafted temporal and textual features of the cascade. We implement two versions of the model as our baseline -- {\bf CasPred (org)} which uses a subset of the original features used, applicable to our setting, and {\bf CasPred (add)} which uses additional features proposed by~\citet{10.1145/3394486.3403251}.

\subsubsection{Neural network baselines} We consider the following three recent neural architectures as baselines:

{\bfseries NNPP} or Neural Network Point Process \citep{neuralnetwork-pointprocess} is an RNN-based method for generalized modeling of temporal point processes.  

{\bfseries DeepHawkes} \cite{10.1145/3132847.3132973} is an end-to-end deep learning framework that combines the predictive power of models based on neural network architectures and interpretability of cascades provided by the Hawkes Process.

{\bfseries DeepCas} \citep{DBLP:conf/www/LiMGM17} is a neural network model for predicting cascade growth. It learns a representation of cascade networks by sampling node sequences through random walks processes, thereby leveraging the structural information of the network.

{\bfseries ChatterNet} \citep{10.1145/3394486.3403251} is a neural network model to predict social chatter intensity leveraging on exogenous and endogenous influence combination. 
To apply it in our setting, we remove the endogenous influence module, resulting in a single LSTM layer integrating exogenous signals from news. Moreover, we incorporate aggregated follower count at each observation bin (similar to \shortname{}) in addition to retweet arrival.

\subsection{Ablation variants}
\label{subsec:ablation}

We seek to investigate the contributions of different components of \framework\ in the overall performance by ablation. We explore the following three ablation variants:

 {\bfseries\framework-text.} We take away the contribution of exogenous influence in this variation by removing the scaled dot-product attention between news and tweet. In this variation, the modulation parameter $\mu$ in Section~\ref{subsec:param_compute} is computed by applying the feed-forward layer transformation on the tweet text representation $X_\tau$ only. 

 {\bfseries\framework-CO.} In this variation, contributions from the tweet content as well as the exogenous influence are ablated; retweet growth parameters are estimated from the cascade growth dynamics in the observation window alone, using the modified LSTM layer. 

 {\bfseries\framework-LSTM.} To investigate the gain in modeling capacity enforced by the modifications we applied on LSTM gates in Section~\ref{subsec:lstm_retweet_arrival}, we replace it with the original LSTM layer with rest of the components unchanged.

\begin{table}[!t]
    \centering
    \caption{Comparison with the baselines and the variants of \framework.  ($\downarrow$: lower value is better). CasPred versions do not predict the actual size of future cascades; hence metrics other than step-$\tau$ are unapplicable for these two baselines. SEISMIC and TiDeH emerge as the best generative baselines in terms of correlation and MAPE, respectively. \shortname{} outperforms the rest of the neural network baselines in both metrics. t/s signifies average inference time per sample.}
    \label{tab:result-overall}
    \scalebox{0.9}{
    \begin{tabular}{l|c|c|c|c|c}
    \hline
        Model & $\tau$ & $\rho$ & MAPE (\%) $\downarrow$ & Step-$\tau$ & t/s (ms.)\\
    \hline
        Hawkes & $0.202$ & $0.277$ & $110.25$ & $0.231$ & $196.72$\\
        SEISMIC & $0.532$ & $0.572$ & $138.86$ & $0.522$ & $67.80$\\
        TiDeH & $0.306$ & $0.403$ & $77.90$ & $0.370$ & $14.59$\\
        NNPP & $0.344$ & $0.427$ & $79.12$ & $0.379$ & $6.23$ \\
        DeepHawkes & $0.315$ & $0.411$ & $71.57$ & $0.326$ & $11.23$\\
        DeepCas & $0.350$ & $0.476$ & $60.69$ & $0.419$ & $9.14$\\
        ChatterNet & $0.342$ & $0.455$ & $63.69$ & $0.404$ & $8.77$\\
        CasPred (org) & - & - & - & $0.231$ & $\mathbf{0.01}$\\
        CasPred (add) & - & - & - & $0.300$ & $0.02$\\\hline
        \framework-LSTM & $0.597$ & $0.769$ & $35.78$ & $0.688$ & $5.54$\\
        \framework-CO & $0.625$ & $0.784$ & $24.16$ & $0.741$ & $1.08$\\
        \framework-text & $0.627$ & $0.789$ & \textbf{24.01} & $0.742$ & $2.19$\\\hline
        \framework & $\textbf{0.633}$ &   $\mathbf{0.793}$ & {25.06} & $\mathbf{0.744}$ & $5.40$\\
    \hline
    \end{tabular}}
\end{table}

\begin{figure}[!t]
    \centering
    \includegraphics[width=\columnwidth]{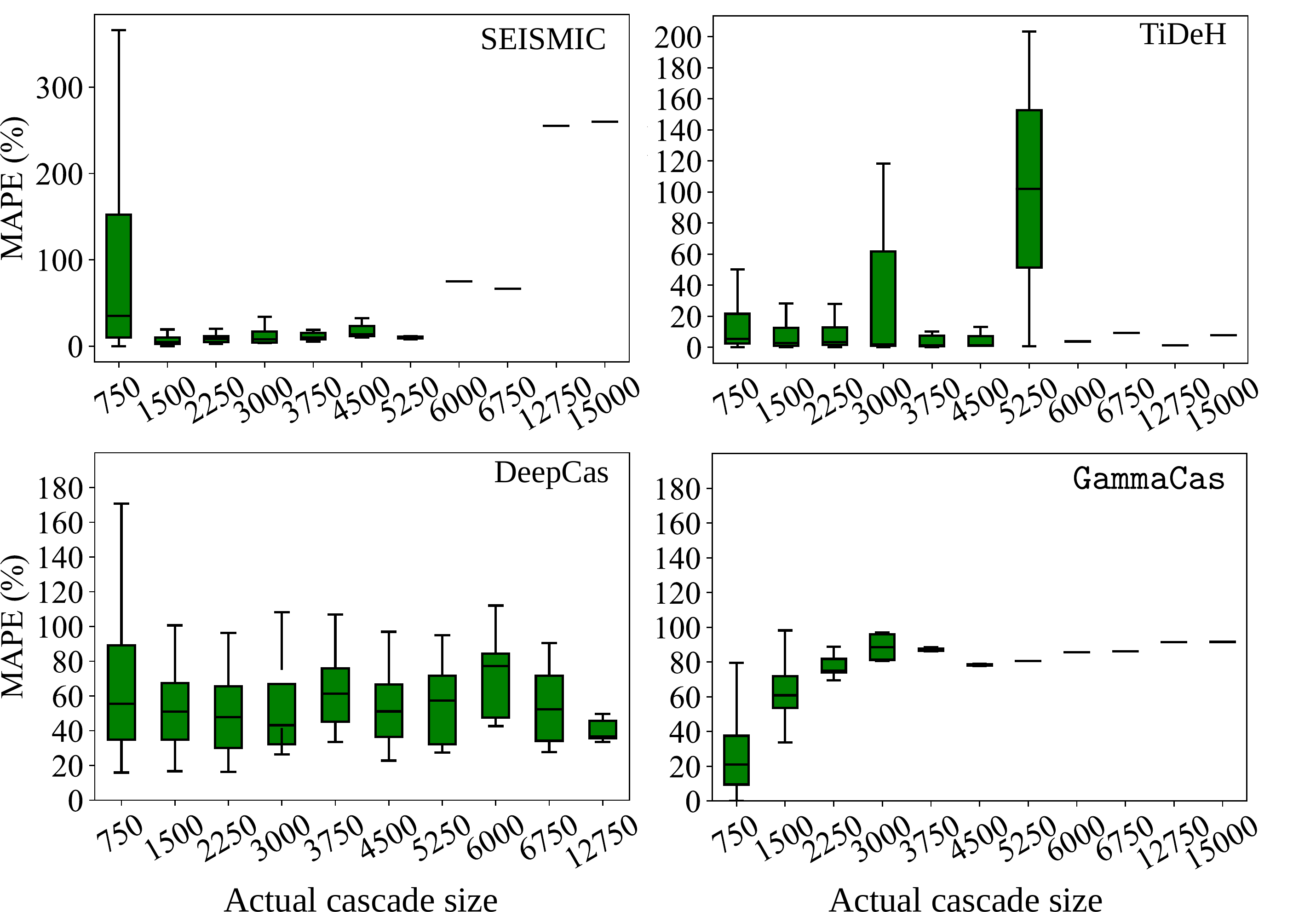}
    \caption{Variation in performance (MAPE) over different cascade sizes for SEISMIC, TiDeH, DeepCas, and \framework. We plot the mean, max, min and standard deviations of absolute percentage error at different bins of cascade sizes.}
    \label{fig:size-MAPE}
\end{figure}
\section{Results and discussion}
\label{sec:results}

The growth of a retweet cascade is a stochastic process that is hard to predict, as random events may shift the growth dynamics of a cascade even after a sufficient observation window. It is important for a model to decide which tweets possess the potential to generate a larger cascade compared to another even when the predicted sizes may not be in range with the actual cascade sizes in a future time. For this reason, we compare \framework, its variants, and all the baselines with three evaluation metrics --- {\bf Mean Absolute Percentage Error (MAPE)} to estimate the difference in predicted and actual sizes; {\bf Kendall's $\tau$} and {\bf Spearman's $\rho$ correlation} between the predicted and actual set of cascade sizes to estimate the models' ability to rank tweets according to their potential to generate cascades. As CasPred predicts whether a cascade will reach a certain size range instead of predicting the actual size, we compute {\bf step-wise Kendall's $\tau$ correlation} \citep{10.1145/3394486.3403251} between the predicted range and the actual range.

\subsection{Overall performance}
\label{subsec:result-overall}

In Table~\ref{tab:result-overall}, we present the performance of \framework, its ablation variants, and baselines to predict cascade size at $24$ hrs. prediction horizon upon $6$ hrs. observation window.

\subsubsection{Comparison among baselines} All the purely generative models (SEISMIC, TiDeH, and Hawkes) yield high MAPE (i.e, poor performance) across all prediction horizons. After investigating the actual predictions made by these three models, we find that these models often overestimate the future cascade size by a large margin (often to an order of $10^3$--$10^4$). Though excluding such cases results in a performance comparable to \framework, the fraction of such overestimating instances is high enough ($>20\%$) to cause performance instability.  {\bf Among the generative baselines, in terms of correlation coefficients, SEISMIC emerges as the best performing generative baseline, while TiDeH stands as best in terms of MAPE.}

\begin{figure}[!t]
    \centering
    \includegraphics[width=\columnwidth]{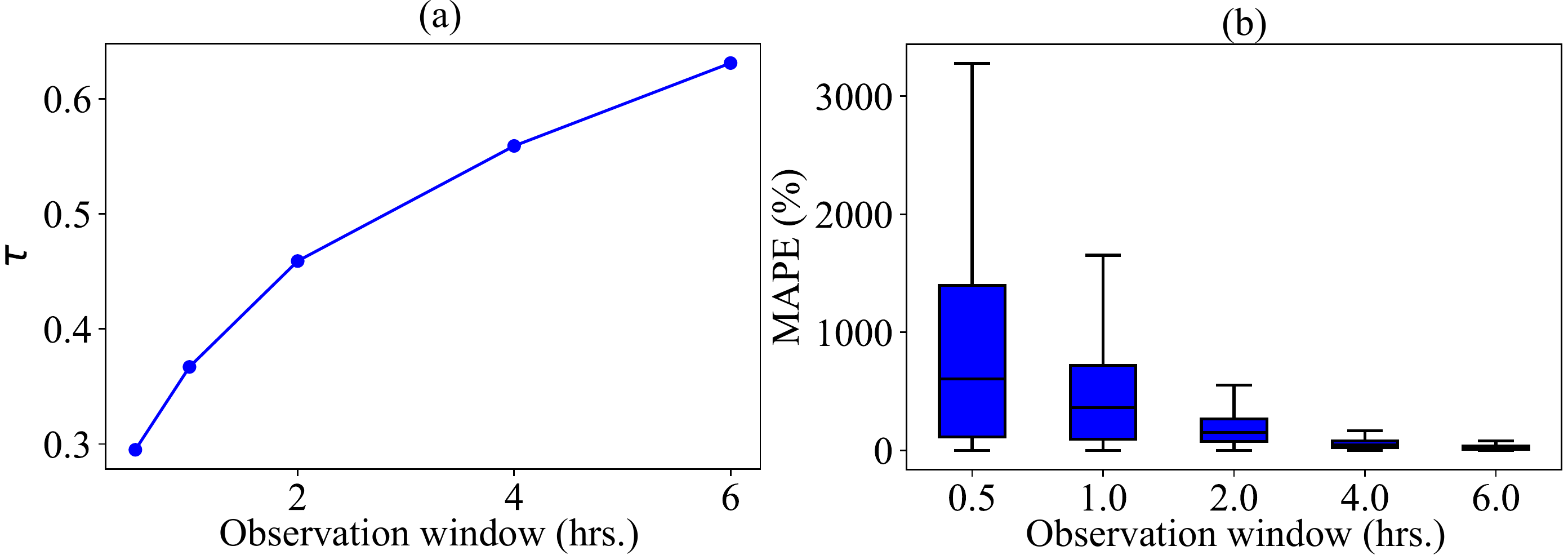}
    \caption{Variation in performance of \framework\ to predict future cascade size at $24$ hours prediction horizon with observation window sizes ($\Delta_{obs}$) $30$ min., $1$ hour, $2$ hours, $4$ hours, and $6$ hours. In (a), we show the correlation in terms of Kendall's $\tau$ between predicted and actual cascade sizes. In (b), we plot the maximum, minimum and mean values of sample-wise absolute percentage errors along with standard deviation.}
    \label{fig:window-vary}
\end{figure}

All the three neural network-based baselines perform closely with respect to all the evaluation metrics, with DeepCas emerging as the best performing one. ChatterNet suffers from the tailoring we had to introduce for the sake of making it applicable to retweet cascade prediction in a different problem setting altogether. Neural network-based model of temporal point processes is able to model cascade growth better compared to simple generative models. However, NNPP does not take any other features except the retweet-arrival statistics. This explains its limitation compared to DeepCas. {\bf In terms of consistent performance on variable-sized cascades and MAPE, we consider DeepCas to be the best performing baseline altogether.}

\subsubsection{Comparing \framework\ with baselines}
From the lowermost block of Table~\ref{tab:result-overall}, it is evident that \framework\ and all its ablation variants perform better than all the baselines by a substantial margin in terms of correlation and absolute error ($\mathbf{18.98\%}$ increase in Kendall's $\tau$ from SEISMIC and $\mathbf{35.63}$ absolute reduction in MAPE compared to DeepCas).  In Figure~\ref{fig:size-MAPE}, we plot how the performance of four highly-ranked competing models, namely SEISMIC, TiDeH, DeepCas, and \framework, are influenced by the actual size of the cascade at $24$ hours prediction horizon. The overshooting problem of SEISMIC and TiDeH is evident from these plots as well. 

All the ablation variants perform closely to \framework; the common signal present in all these models is the temporal dynamics of retweet arrival within the observation window. One may trivially decide this to be the most important signal for modeling cascade growth dynamics. 
However, we can observe significant improvement of correlation measures once we introduce the exogenous influence-modulated signals. Interestingly, the overall MAPE error decreases slightly with some ablated variants. We investigate the influence of tweet content and exogenous signals later in Section~\ref{subsec:diagnostics} while diagnosing \framework\ predictions.

The design choice we made to introduce extra gating mechanism to LSTM cell to model retweet arrival dynamics evidently brings performance gain. As seen in Table~\ref{tab:result-overall}, \framework-LSTM (with all signals included) is outperformed \framework\ as well as rest of the ablation variants.

We also investigate the latency of prediction for all the models in Table~\ref{tab:result-overall}. Generative models usually take longer to predict per sample as they use the observation window to estimate the parameters using a likelihood measure. As CasPred solely depends on a manually engineered feature set and does not need any temporal processing (thereby reducing the number of operations), it emerges as the fastest inferring model. Among the rest, \framework\ is an order of magnitude faster than the models which show comparable accuracy. 
As expected, ablated variants with no news-tweet attention or textual features are faster than full \framework.

\subsection{Variation with observation window}
\label{subsec:vary-obs}

\begin{figure}[!t]
    \centering
    \includegraphics[width=\columnwidth]{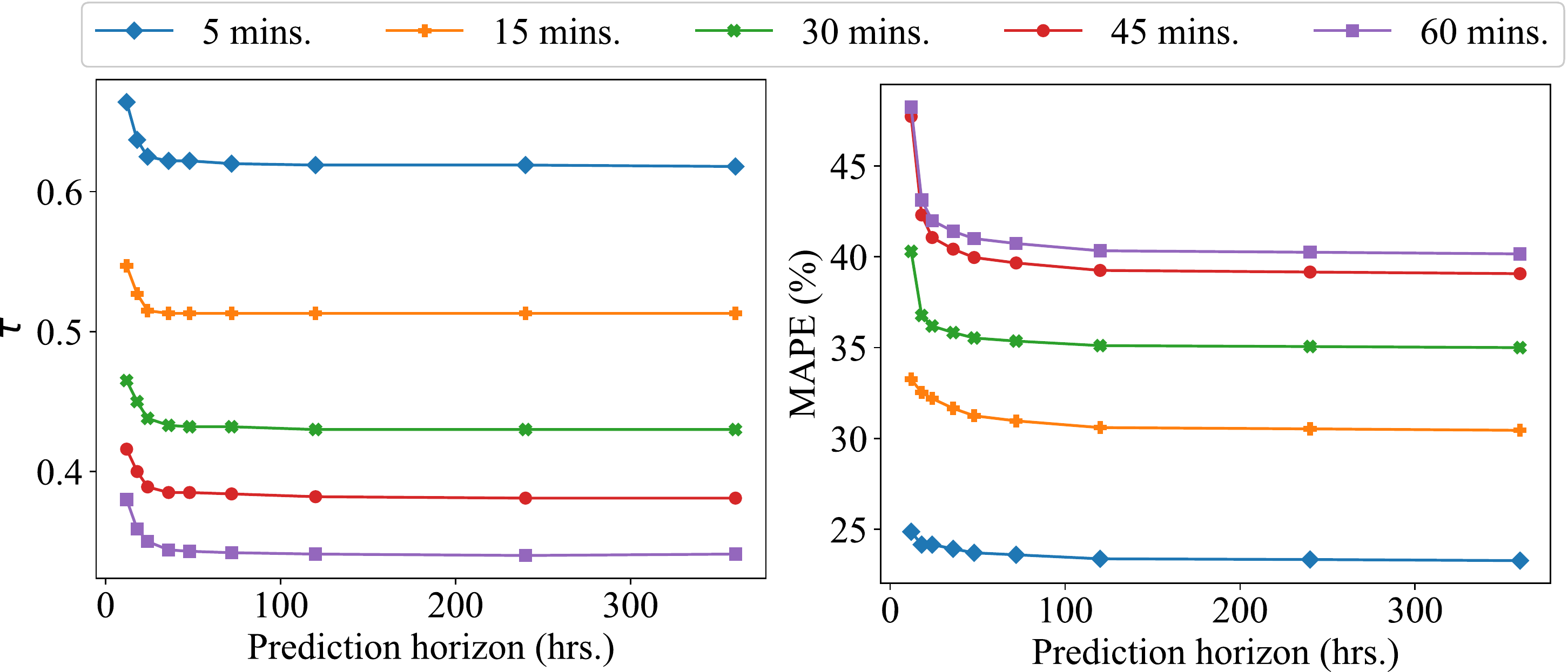}
    \vspace{-5mm}
    \caption{Variation in performance of \framework\ at different prediction horizons ($\Delta_p$) for different widths of observation bins ($\Delta_o$). We evaluate this performance in terms of Kendall's $\tau$ and MAPE. With coarser binning (larger $\Delta_o$), the performance drops significantly. }
    \label{fig:bin-variation}
\end{figure}

As past studies suggested \cite{10.1145/2566486.2567997}, a sufficient amount of early observation is necessary to estimate the future size of a cascade. \framework\ offers the flexibility of using different observation windows due to its temporal processing of the input along with an incremental estimation of the growth parameters. In Figure~\ref{fig:window-vary}, we show the variation of performance of \framework\ for multiple observation windows. Evidently, a larger observation window helps predict the future cascade size with better accuracy. However, even with a shorter observation window ($4$ hours), \framework\ outperforms all the baseline models in terms of correlation and absolute percentage error. 

Splitting the cascade dynamics within the observation window into successive bins of retweet arrival and aggregate follower counts serves as a uniform discretization of the irregular arrival processes. 
Intuitively, a smaller temporal bin width would result in a more accurate approximation of time, leading to superior performance. This is also evident in Figure~\ref{fig:bin-variation}, where we plot $\tau$ (left) and MAPE (right) of \framework\ for predicting cascade sizes at different prediction horizons when using different bin widths ($5, 15, 30, 45$ and $60$ mins.). While with narrower bins, the performance drop from near to distant prediction horizons is steep, it effectively flattens with the higher error rate in longer bins. However, narrow bins result in a longer sequence of input, resulting in longer recurrence relations to be captured and higher training/testing cost.

\subsection{Variation with prediction horizons}
\label{subsec:vary-pred-horizon}
\begin{figure}[!t]
    \centering
    \includegraphics[width=\columnwidth]{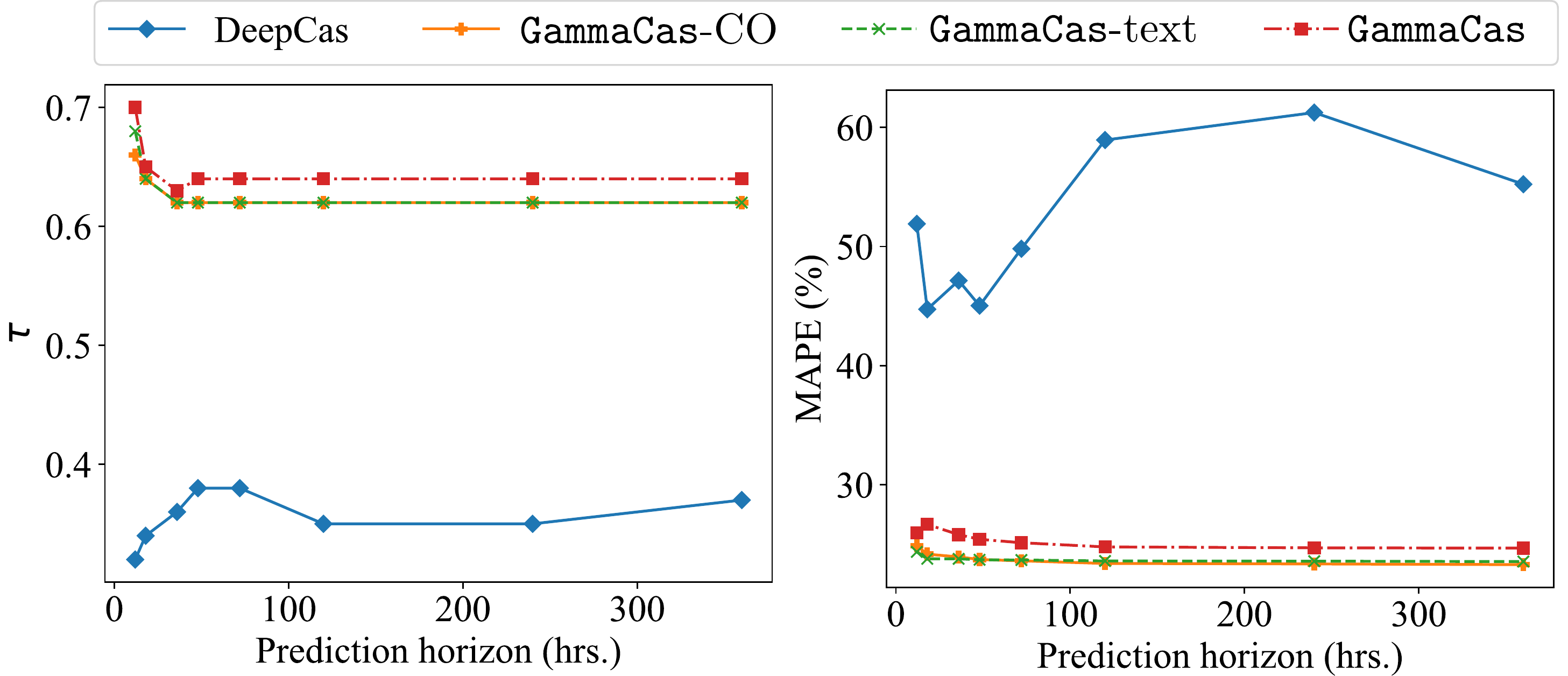}
    \vspace{-5mm}
    \caption{Variation of performance of \framework, its ablation variants, and the best performing baseline DeepCas, on different prediction horizons.}
    \label{fig:pred-variation}
\end{figure} 
The quality of fit for the estimated parameters of a monotone function of time is judged by how they fit at different future horizons. We vary the prediction horizon and observe the evaluations for \framework, its ablation variants, and the best-performing baseline, DeepCas.  As shown in Figure~\ref{fig:pred-variation}, \framework\, and its ablation variants produce a more stable performance over different horizons, compared to DeepCas. While in terms of correlation, \framework\ shows an initial performance drop as the prediction horizon increases, we can see an almost consistent MAPE over all the horizons. Moreover, models like DeepCas need to be trained and tested for each prediction horizon separately, while \framework\ offers a flexible prediction setting much similar to its generative counterparts, adding significance to the judgment of parameter utility.

\subsection{Diagnostic experiments on \framework}
\label{subsec:diagnostics}

\begin{figure}[!t]
    \centering
    \includegraphics[width=\columnwidth]{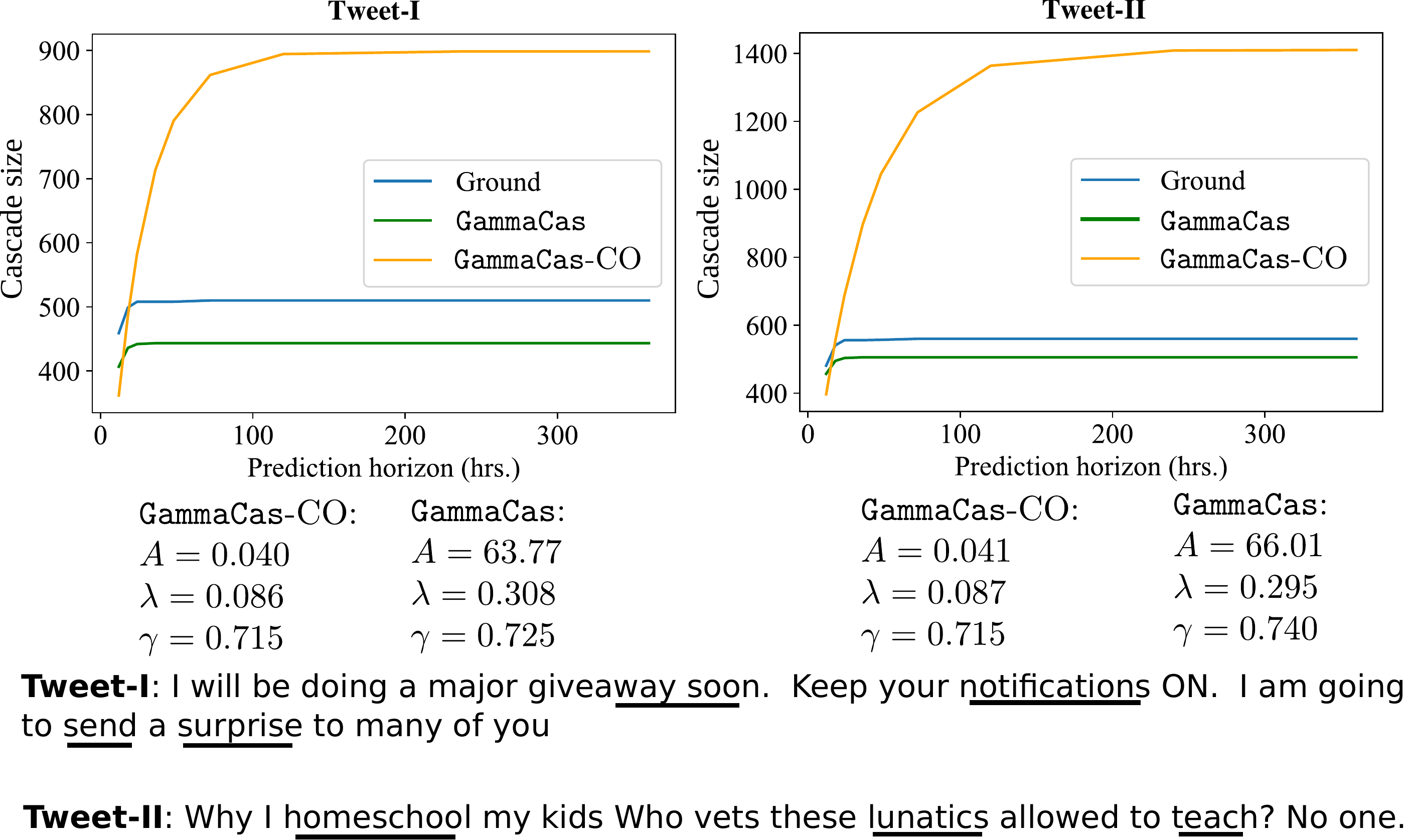}
    \caption{Predicted and actual cascade sizes for two tweets by \framework\ and \framework-CO. Underlined words in the tweets are those attaining higher attention weights. In both the cases, \framework-CO meets a very low value of $\lambda$ entangled with a low value of $A$, which leads to overshooting the cascade size.}
    \label{fig:case-study}
    \vspace{-5mm}
\end{figure}

\begin{figure}
    \centering
    \includegraphics[width=\columnwidth]{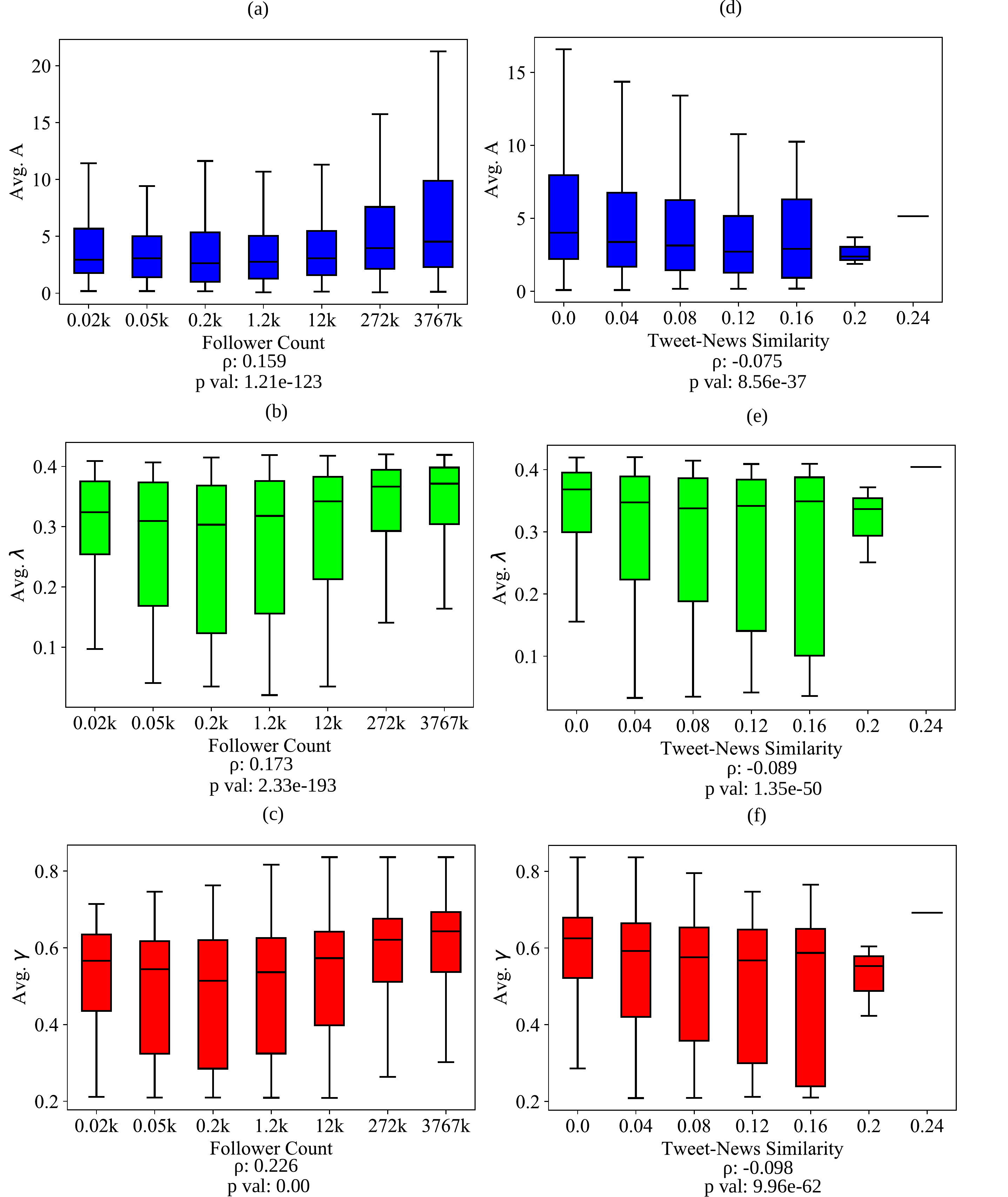}
    \caption{Variations of $A$, $\lambda$, and $\gamma$ estimated by \framework\ with follower count of the root user and news-tweet similarity. We plot the max, min, mean and standard deviation of the parameters for tweets at different bins of follower count/news-tweet similarity. We also show the correlations between each pair of variables in terms of Spearman's $\rho$ and the corresponding $p$-value.}
    \vspace{-5mm}
    \label{fig:parameter-plot}
\end{figure}
 
In Section~\ref{sec:model}, we provided intuitive justifications of our design decisions. To look for the potential presence of more profound connections between different influencing signals and the cascade growth parameters that \framework\ attempts to model, we look into individual predictions as well as the overall distribution of parameters. 

In Figure~\ref{fig:case-study}, we present two example tweets, actual sizes of the cascades they generate, and the predicted sizes by \framework\ and \framework-CO over different prediction horizons. While {\bf Tweet-I} was from a popular social media influencer addressing their fan-base (no exogenous influence), {\bf Tweet-II} was regarding a teacher passing abusive remarks towards students in the context of COVID-19 (triggered by exogenous event). In both cases, \framework-CO, in the absence of content-based signals, underestimates $A$ and to fit the observed retweet arrivals, underestimates the decay parameter $\lambda$ as well. This leads to overshooting the actual cascade size by a large margin. The low value of $\lambda$ also sets a longer supercritical phase of the cascades. On the other hand, \framework\ estimates a much higher value of $A$ with larger $\lambda$ decay, providing a better approximation of the future cascade size.

We extract the attention values $\alpha_i$ (see Equation~\ref{eq:text_attention}) for each token (other than stopwords) of the tweets. In Figure~\ref{fig:case-study}, we mark the words receiving significant attention. It is evident that certain topic-signaling and positive/negative sentiment words put a higher contribution constituting the signals deciding cascade growth.

To investigate the effects of follower count of the root users and exogenous influence on the cascade growth parameters estimated by \framework, we plot one-to-one mappings between them in Figure~\ref{fig:parameter-plot}. We compute the correlations between each pair of variables to find out their statistical significance. Evidently, the follower count of the root user holds a strong influence on all of the three parameters (subplots (a), (b), and (c) in Figure~\ref{fig:parameter-plot}). However, the growth parameter $\gamma$ is the most positively correlated one. Intuitively, one can translate this as high follower count ensures an influential user with a high degree of organic reach; when such a user tweets something, the rate of growth at the supercritical stage is likely to be higher compared to some less influential user. Alternatively, if the root user of the cascade reaches a large number of users directly, the subsequent levels are likely to have a lower value of average out-degree and thereby, decreasing the rate of subsequent cascade growth. This points to the high value of the decay parameter $\lambda$ as well. Lastly, users become influential with historical activity, i.e., the degree of diffusion of contents posted by them are usually high, pointing towards a possible positive reinforcement of $A$ in the future cascades they cause.

As opposed to the follower count, similarity of a tweet with news articles published in the past $6$ hrs. shows a weakly negative (yet statistically significant) correlation with all three of the parameters. In this case, the effect is strongest in the case of both $\gamma$ and $\lambda$, pointing towards a slow growth as well as decay when the similarity is high, and vice versa. This weakly negative correlation is consistent with our findings shown in Figure~\ref{fig:intro_fig}(b) in Section~\ref{sec:Intro}, where we observed a similar weakly negative impact of similarity between a tweet and past news on the cascade size. Again, a plausible intuition behind this might be that the potential of a tweet be the genesis of a large cascade is facilitated if it brings new, hitherto unknown information.

\section{Conclusion and Future Work}
\label{sec:End}

We presented \shortname, a new deep cascade prediction architecture that combines content, network, and exogenous signals into a transparent, parameterized time integral.  Prediction loss can be back-propagated to the feature-processing networks. We prepared a large-scale dataset of retweet cascades and time-aligned news texts, and provided insightful findings on the dynamics of cascade growth.  \shortname{} provides a better and more robust cascade size prediction compared to recent competitive baselines on different prediction horizons with varying early observation window. Investigations on parametric functions and feature representations learned by \shortname{} provide a meaningful interpretation of relations between cascade dynamics and various input features related to exogenous influences obtained from online news articles, the textual content of a tweet, degree distribution of cascade participant nodes, retweet arrival, etc.

As a future extension, one may intend to introduce multimodal signals introduced by richer metadata of the tweet (images, memes, videos, etc.). Information cascades formed from a tweet are not limited to simple retweet trees as well. For example, link to an existing tweet may be posted as standalone tweets. When such a tweet gets retweeted, this practically forms an extended information cascade of the original tweet. These complex dynamics makes the cascade modeling problem intrinsically challenging. Moreover, information cascades in general goes beyond the resharing (via retweet, quote, or links) mechanism. A certain news (and specially in the present day scenario, a fake one) may form cascades of diffusion via independent tweets. Modeling such dynamics using the various signals we used is likely to provide further insights.

{\small
\bibliographystyle{IEEEtranN}
\bibliography{ref-new}}

\IEEEdisplaynontitleabstractindextext

\IEEEpeerreviewmaketitle

\vspace{-5mm}

\begin{IEEEbiography}[{\includegraphics[width=1in,height=1.25in,clip,keepaspectratio]{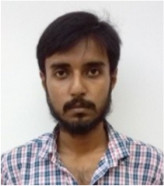}}]{Subhabrata Dutta} is a PhD student in the Department of Computer Science and Engineering, Jadavpur University. His research interests include Social Computing, Natural Language Processing, and Machine Learning. 

\end{IEEEbiography}
\vspace{-5mm}

\begin{IEEEbiography}[{\includegraphics[width=1in,height=1.25in,clip,keepaspectratio]{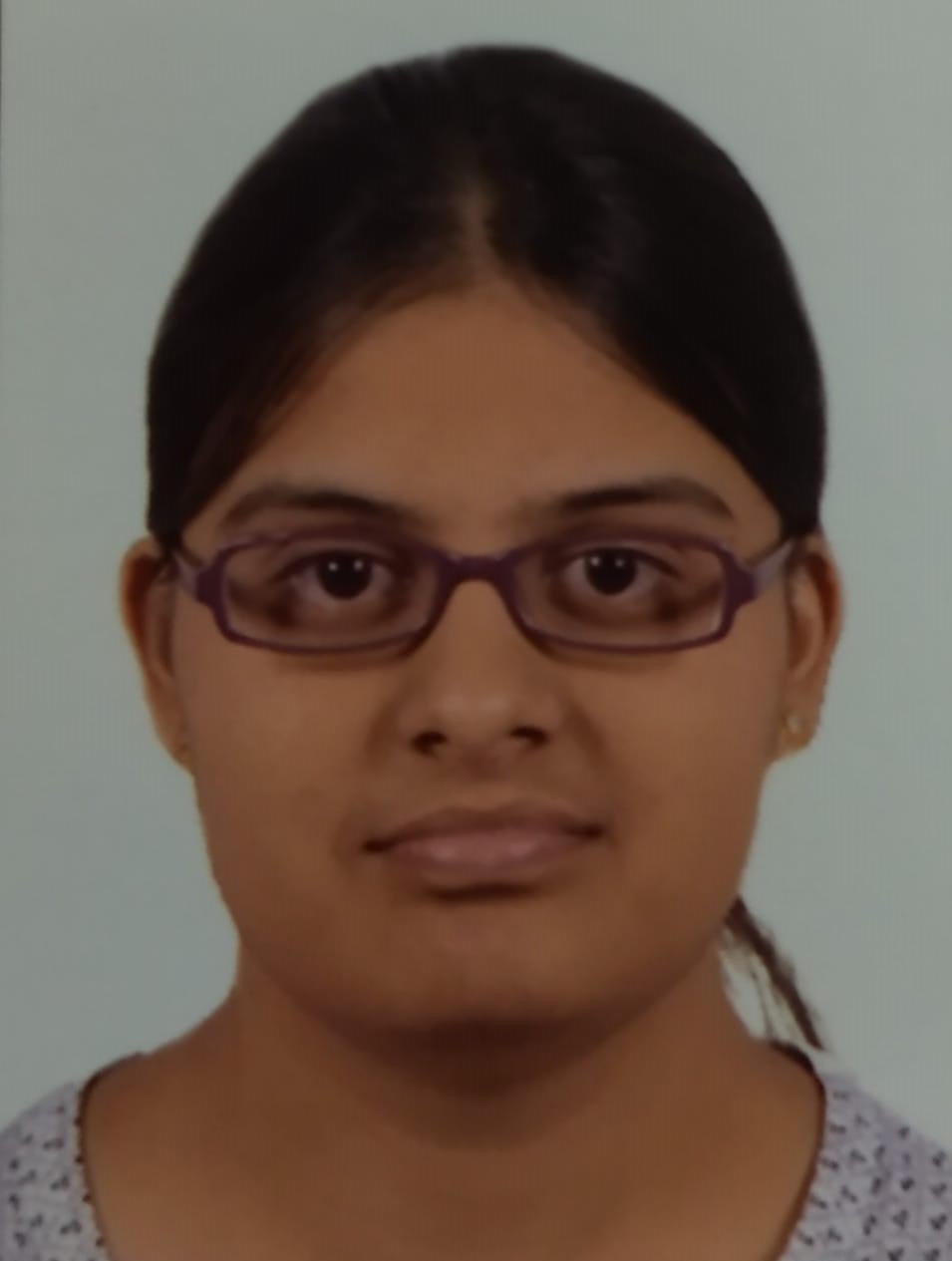}}]{Shravika Mittal} is a Software Development Engineer at Adobe, India. She completed her undergraduate studies in Computer Science and Engineering from IIIT-Delhi. Her research interests include Social Network Analysis, Network Science, and Natural Language Processing. She received the Chancellor's gold medal, Dean's list for Innovation in Research and Development as an undergraduate student.
\end{IEEEbiography}
\vspace{-5mm}

\begin{IEEEbiography}[{\includegraphics[width=1in,height=1.25in,clip,keepaspectratio]{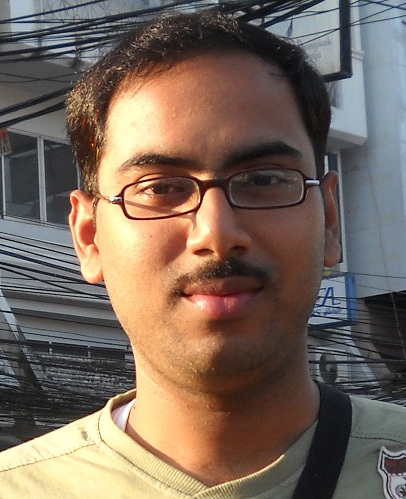}}]{Dipankar Das} is an Assistant Professor in the Department of Computer Science and Engineering, Jadavpur University and Visveswaraya Young Faculty, Ministry of Electronics and Information Technology (MeitY), Government of India. He is presently leading four research projects of DRDO, SERB, DST, UGC, Govt. of India. His research interests are in the area of Natural Language Processing / Computational Linguistics, Emotion and Sentiment Analysis, Search Engine and Information Extraction, Machine Learning, Deep Learning, Social Network Analysis, Data Science (Big Data) and so on.

\end{IEEEbiography}
\vspace{-5mm}

\begin{IEEEbiography}[{\includegraphics[width=1in,height=1.25in,clip,keepaspectratio]{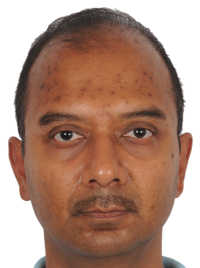}}]{Soumen Chakrabarti}
is a Professor of Computer Science at IIT Bombay. He works on knowledge graphs, question answering, and social networks. He has published extensively in WWW, SIGKDD, ACL, EMNLP, IJCAI, AAAI, VLDB, SIGIR, ICDE and other conferences. He won the best paper award at WWW 1999. He was coauthor on the best student paper at ECML 2008. His work on keyword search in databases got the 10-year influential paper award at ICDE 2012. He got his PhD from University of California, Berkeley and worked at IBM Almaden, CMU and Google in the past.  He received the Bhatnagar Prize in 2014 and the Jagadis Bose Fellowship in 2019.
\end{IEEEbiography}
 
\vspace{-5mm}

\begin{IEEEbiography}[{\includegraphics[width=1in,height=1.25in,clip,keepaspectratio]{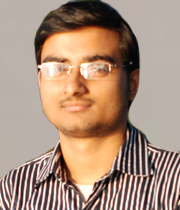}}]{Tanmoy Chakraborty}
is an Assistant Professor and a Ramanujan Fellow at the Dept. of CSE, IIIT-Delhi, India, where he leads a research group,  Laboratory for Computational Social Systems (LCS2). His primary research interests include Social Computing and Natural Language Processing. He has received several awards including  Faculty Awards from Google, IBM and Accenture; Early Career Research Award, DAAD Faculty Fellowship. He is a member of ACM and IEEE. More details at \url{http://faculty.iiitd.ac.in/~tanmoy/}.
\end{IEEEbiography}

\end{document}